\documentclass[english,aps,preprint]{revtex4}
\usepackage[]{fontenc}
\usepackage[latin9]{inputenc}
\usepackage{amsmath}
\usepackage{amssymb}
\usepackage{graphicx}
\usepackage{esint}

\makeatletter
\@ifundefined{textcolor}{}
{%
 \definecolor{BLACK}{gray}{0}
 \definecolor{WHITE}{gray}{1}
 \definecolor{RED}{rgb}{1,0,0}
 \definecolor{GREEN}{rgb}{0,1,0}
 \definecolor{BLUE}{rgb}{0,0,1}
 \definecolor{CYAN}{cmyk}{1,0,0,0}
 \definecolor{MAGENTA}{cmyk}{0,1,0,0}
 \definecolor{YELLOW}{cmyk}{0,0,1,0}
}

\usepackage{slashed}

\makeatother

\usepackage{babel}
\begin{document}

\title{Two-loop finiteness of self-energies in higher-derivative SQED$_{3}$}

\author{E. A. Gallegos}

\email{egallegoscollado@gmail.com}

\selectlanguage{english}%

\affiliation{Departamento de Física, Universidade Federal de Santa Catarina, Campus
Trindade, 88040-900, Florianópolis, SC, Brazil}

\author{R. Baptista}

\email{raybap@gmail.com}

\selectlanguage{english}%

\affiliation{Departamento de Física, Universidade Federal de Santa Catarina, Campus
Trindade, 88040-900, Florianópolis, SC, Brazil}
\begin{abstract}
In the $\mathcal{N}=1$ superfield formalism, two higher-derivative
kinetic operators (Lee-Wick operators) are implemented into the standard
three dimensional supersymmetric quantum electrodynamics (SQED$_{3}$)
for improving its ultraviolet behavior. It is shown in particular
that the ghosts associated with these Lee-Wick operators allow to
remove all ultraviolet divergences in the scalar and gauge self-energies
at two-loop level. 
\end{abstract}
\maketitle

\section{INTRODUCTION\label{sec:Sec1}}

Higher-derivative supersymmetric theories bring together two entirely
distinct mechanisms of removing ultraviolet (UV) divergences in the
perturbation formulation. In higher-derivative theories the UV divergences
are removed by an exchange of positive- and negative-metric (ghosts)
states, while in supersymmetric theories the UV cancellation occurs
by an exchange of virtual particles with opposite statistics. The
ghost stumbling blocks (lost of unitarity and Lorentz violation),
on the other hand, can be avoided for instance by applying the Lee-Wick's
ideas \cite{Lee-Wick (1969)} and by adopting the Feynman contour
prescription of Cutkosky et al. \cite{Cutkosky (1969)}.

One of us and coworkers \cite{Gall-Sen-Silv (2013)} has recently
investigated the classical and quantum effects of higher derivative
operators in the three dimensional Wess-Zumino (WZ) model within the
superfield approach. In that work, aside from the study of the one-loop
effective potential, it was shown that the two-loop scalar self-energy
becomes to be finite by introducing a single higher-derivative operator
in the kinetic part of the usual WZ action. Since the analytical continuation
from the Minkowski space to the Euclidean space is lost in higher-derivative
theories due to the presence of complex poles in the complex energy
plane, the finiteness of the two-loop self-energy of the Wess-Zumino
model rests on the assumption that all residues of the complex poles
involved in the Cauchy's theorem for performing the ``Wick rotation''
are finite. As argued in \cite{Gall-Sen-Silv (2013)}, the assumption
about the finiteness of the residues, just like at one-loop level
in four dimensions \cite{Grinstein & O'Connell (2008)}, is valid
at two-loop level in three dimensions. Therefore as far as the analysis
of ultraviolet divergences is concerned the choice of Feynman contours
in momentum integrals is irrelevant.

In this paper we continue with our study of higher-derivative operators
in the $\mathcal{N}=1$ superfield approach at three dimensions. Specifically,
we deform the standard supersymmetric quantum electrodynamics (SQED$_{3}$)
with scalar self-interaction by implementing two higher-derivative
operators in its kinetic action. One of them is the susy extension
of the Lee-Wick operator $\partial_{\mu}F^{\mu\nu}\partial_{\rho}F_{\:\:\nu}^{\rho}$
and the other is the gauge extension of the Lee-Wick-WZ operator $D^{2}\bar{\Phi}D^{2}\Phi$.
We show that the implementation of these two higher-derivative operators
in standard SQED$_{3}$ is enough to remove all residual susy divergences
in the scalar and gauge self-energies at two-loop level. 

Our paper is planned as follows. In Sec. \ref{sec:Sec2} we discuss
in general terms the construction of supersymmetric gauge theories
in the $\mathcal{N}=1$ superfield formalism at three dimensions by
imposing the symmetry and renormalization requirements. It is shown
that the most general renormalizable superspace Lagrangian is set
up of operators with mass dimension less than or equal to 2. We emphasize
also that this result depends on the definition of the kinetic part
of the action which describes the theory. To close this section, we
focus our attention on the SQED$_{3}$ with scalar self-coupling and
identify its power counting divergences in the scalar and gauge self-energies
at two-loop order. In Sec. \ref{sec:Sec3}, as commented above, we
implement two higher-derivative operators in SQED$_{3}$. These operators
on incorporating in the kinetic part of the SQED$_{3}$ action improve
the convergence of the propagators in such a way that all usual two-loop
divergences in the scalar and gauge self-energies are removed. Our
results are summarized in Sec. \ref{sec:Sec4}.

\section{$\mathcal{N}=1$ supersymmetric gauge theories in three dimensions
\label{sec:Sec2}}

In this section we construct the most general (lower-derivative) Abelian
gauge theory in the $\mathcal{N}=1$ superfield formalism at three
dimensions by using the symmetry and renormalization requirements.
As in the next section we shall extend the standard SQED$_{3}$ by
introducing higher-derivative operators in its action, we emphasize
the importance of the definition of the kinetic part of the action
in the renormalization procedure. Finally, we classify the sorts of
UV divergences which are present in the standard SQED$_{3}$ with
scalar self-coupling. 

Concretely, we will construct a theory (with lower-derivatives) which
describes the evolution of the superfields
\begin{equation}
\Phi\left(z\right)=\varphi\left(x\right)+\theta^{\alpha}\psi_{\alpha}\left(x\right)-\theta^{2}F\left(x\right),\label{eq:2.1}
\end{equation}
and
\begin{equation}
A_{\alpha}\left(z\right)=\chi_{\alpha}\left(x\right)-\theta_{\alpha}B\left(x\right)+i\theta^{\beta}V_{\alpha\beta}\left(x\right)+\theta^{2}\left(-i\partial_{\alpha\beta}\chi^{\beta}-2\lambda_{\alpha}\right).\label{eq:2.2}
\end{equation}
Here $\Phi\left(z\right)$ and $A_{\alpha}\left(z\right)$ stand for
a scalar superfield and a spinor superfield, respectively. In this
paper we adopt the notation of \cite{Gates-etal}. 

This theory in addition has to be supersymmetric, Lorentz invariant
and to satisfy the following (gauge) symmetry:
\begin{equation}
\Phi'\left(z\right)=\mbox{e}^{-ieK}\Phi\left(z\right)\qquad A'_{\alpha}\left(z\right)=A_{\alpha}\left(z\right)+D_{\alpha}K\left(z\right),\label{eq:2.3}
\end{equation}
where $D_{\alpha}\left(=\mbox{\ensuremath{\partial}}_{\alpha}+i\theta^{\beta}\partial_{\alpha\beta}\right)$
is the susy-covariant derivative and $K\left(z\right)$ denotes an
arbitrary scalar superfield. 

In order to be able to make use of the renormalization requirement,
we will need to compute the superficial degree of divergence and to
this end we must first define the kinetic part of action. 

For reasons to be explained below, we start with the action 
\begin{equation}
S=\int d^{5}z\left\{ \frac{1}{2}W^{\alpha}W_{\alpha}+\bar{\Phi}D^{2}\Phi+U\left(A_{\alpha},\Phi\right)\right\} ,\label{eq:2.4}
\end{equation}
where $W_{\alpha}$ is the field strength and is defined as $W_{\alpha}=\frac{1}{2}D^{\beta}D_{\alpha}A_{\beta}$.
By using the identity $D^{\alpha}D_{\beta}D_{\alpha}=0$, it is easy
to prove that $W_{\alpha}$ is invariant under the gauge transformations
(\ref{eq:2.3}). $U\left(A_{\alpha},\Phi\right)$, on the other hand,
denotes all the other operators (vertices) which are compatible with
the symmetry and renormalization requirements. To respect the supersymmetry,
$U\left(A_{\alpha},\Phi\right)$ must be constructed from covariant
objects (superfields and susy-covariant derivatives), and so its most
general form has to be
\begin{equation}
U\sim\left(D_{\alpha}\right)^{N_{D}}\left(A_{\alpha}\right)^{N_{A}}\left(\bar{\Phi}\Phi\right)^{N_{\Phi}/2},\label{eq:2.5}
\end{equation}
where $N_{D}$ represents the number of spinor susy-covariant derivatives,
$N_{A}$ the number of spinor superfields, and $N_{\Phi}$ the number
of scalar superfields. Even though $i\partial_{\alpha\beta}$ is also
a (spacetime) susy-covariant derivative, it has not been included
in (\ref{eq:2.5}) because it can be eliminated by means of the identity
$i\partial_{\alpha\beta}=\frac{1}{2}\left\{ D_{\alpha},D_{\beta}\right\} $.
Note also that the symmetries of the theory restrict strongly the
values of $N_{D}$, $N_{A}$ and $N_{\Phi}$. In particular, rotational
symmetry implies that $N_{\Phi}$ must take only even values ($N_{\Phi}=0,\,2,\,\cdots$),
while Lorentz symmetry demands a complete spinor contraction as well
as that $N_{D}+N_{A}=\mbox{even number}$.

Before computing the propagators of the theory and analyzing its renormalization,
some comments are in order with respect to the action (\ref{eq:2.4}).
First of all, it should be noted that one only needs to know the asymptotic
behavior of the propagators (i.e. their behavior when $k^{2}\rightarrow\infty$)
for determining the superficial degree of divergence. Since these
propagators are obtained by inverting the kernels of the quadratic
parts in the fields, they will depend on the definition of the kinetic
part of the action. The most general kinetic action (without considering
the gauge symmetry) has the form
\begin{equation}
S_{0}=\int d^{5}z\left\{ \frac{1}{2}A^{\alpha}\mathcal{O}_{\alpha\beta}A^{\beta}+\bar{\Phi}\mathcal{O}\Phi\right\} ,
\end{equation}
where
\begin{eqnarray}
\mathcal{O}_{\alpha\beta} & = & \sum_{i=0}^{1}\left[r_{i}R_{i,\alpha\beta}+s_{i}S_{i,\alpha\beta}\right]\nonumber \\
 & = & \left(r_{0}+r_{1}D^{2}\right)i\partial_{\alpha\beta}+\left(s_{0}+s_{1}D^{2}\right)C_{\alpha\beta}\\
\mathcal{O} & = & c_{0}+c_{1}D^{2}.
\end{eqnarray}
Here $R_{i,\alpha\beta}$ and $S_{i,\alpha\beta}$ constitute a basis
in the gauge sector \cite{Boldo-Gallegos}, while $r_{i}$, $s_{i}$,
and $c_{i}$ are functions of the d'Alembertian operator $\square=\partial^{\mu}\partial_{\mu}$.
Clearly, if one wants a local theory these functions have to be polynomials
in $\square$. We have chosen (\ref{eq:2.4}) as the starting action
because the gauge invariant operator $W^{\alpha}W_{\alpha}$ contains
after carrying out the integration over $\theta$ the Maxwell Lagrangian,
\begin{equation}
\left.D^{2}\left(W^{\alpha}W_{\alpha}\right)\right|=-\frac{1}{8}F^{\mu\nu}F_{\mu\nu}+\cdots,
\end{equation}
where $F_{\mu\nu}=\partial_{\mu}v_{\nu}-\partial_{\nu}v_{\mu}$. Second,
since $\bar{\Phi}D^{2}\Phi$ is not gauge invariant, other operators
must be included in the action (\ref{eq:2.4}) to restore its gauge
invariance. In fact these operators are encoded in $U$ and will be
found by means of the renormalization condition.

As in conventional gauge theories, to determine the propagator of
the gauge superfield it is necessary to fix the gauge (i.e. to eliminate
the redundant degrees of freedom). For simplicity, we choose the gauge
fixing term
\begin{equation}
S_{gf}=-\frac{1}{4\alpha}\int d^{5}zD^{\alpha}A_{\alpha}D^{2}D^{\beta}A_{\beta}.\label{eq:2.6}
\end{equation}
This gauge fixing term has the advantage of uncoupling the physical
superfields from the (anti-)ghost superfields in the perturbation
approach.

Inserting (\ref{eq:2.6}) into (\ref{eq:2.4}), we easily calculate
with the help of the techniques described in \cite{Boldo-Gallegos}
the propagators for the scalar and gauge superfields. They are given
by
\begin{equation}
\left\langle \bar{\Phi}\left(k,\,\theta\right)\Phi\left(-k,\,\theta\right)\right\rangle =-i\frac{D^{2}}{k^{2}}\delta^{2}\left(\theta-\theta'\right)\label{eq:2.7a}
\end{equation}
and 
\begin{equation}
\left\langle A_{\alpha}\left(k,\,\theta\right)A_{\beta}\left(-k,\,\theta\right)\right\rangle =-\frac{i}{2k^{2}}\left\{ \left(\alpha+1\right)C_{\alpha\beta}+\frac{\left(\alpha-1\right)}{k^{2}}k_{\alpha\beta}D^{2}\right\} \delta^{2}\left(\theta-\theta'\right).\label{eq:2.7b}
\end{equation}

By virtue of these results, the superficial degree of divergence $\omega$
for any arbitrary Feynman diagram of the model reads 
\begin{equation}
\omega=I\, V+2-\frac{E_{\Phi}}{2}-\frac{n_{D}}{2},\label{eq:2.8a}
\end{equation}
where $I$ represents the index of divergence corresponding to the
generic vertex (\ref{eq:2.5}) and is given by
\begin{equation}
I=\frac{N_{\Phi}+N_{D}}{2}-2,\label{eq:2.8b}
\end{equation}
$V$ is the number of vertices, $E_{\Phi}$ is the number of external
$\Phi$ lines, and $n_{D}$ is the number of susy-covariant derivatives
($D_{\alpha}$) transferred to the external lines during the Grassmann
reduction procedure (D-algebra) \cite{Gates-etal}.

From (\ref{eq:2.8a}) we can see that our theory will be renormalizable
if and only if the condition $I\leq0$ is satisfied. That is, if $N_{\Phi}+N_{D}\leq4$.
Despite the fact that $I$ does not depend explicitly on $N_{A}$
so that any vertex with $I\leq0$ regardless of the gauge lines is
acceptable, we shall see below that the symmetries of the theory limit
strongly the form of the vertices, in particular, the number of the
gauge lines.

Consider first the gauge sector, i.e., set $N_{\Phi}=0$ in (\ref{eq:2.8b}).
This condition implies that any renormalizable operator has to be
made up of at most four susy-covariant derivatives ($N_{D}\leq4$).

Using the Lorentz condition $\left(N_{D}+N_{A}=\mbox{even}\right)$
and the gauge symmetry, it is possible to show that there are only
two gauge operators which obey all physical requirements. They are
given by
\begin{equation}
D^{\beta}D^{\alpha}A_{\beta}A_{\alpha}\sim W^{\alpha}A_{\alpha}\,\,\left(N_{D}=2\right)\qquad\: D^{\beta}D^{\alpha}A_{\beta}D^{\gamma}D_{\alpha}A_{\gamma}\sim W^{\alpha}W_{\alpha}\,\,\left(N_{D}=4\right).\label{eq:2.9}
\end{equation}
The former is gauge invariant only within the superspace integral
and gives rise after performing the Grassmann integral to the well-known
Chern-Simons term,
\begin{equation}
\left.D^{2}\left(W^{\alpha}A_{\alpha}\right)\right|=-\frac{1}{2}\varepsilon^{\mu\nu\rho}v_{\mu}\partial_{\nu}v_{\rho}+\cdots,
\end{equation}
whereas the latter is a bona fide operator in the sense that is gauge
invariant out of the superspace integral. We stress that any other
combination of gauge fields and susy-covariant derivatives is forbidden
either by renormalization or by gauge invariance. For instance, an
operator of the fashion $W^{\alpha}A_{\alpha}W^{\beta}A_{\beta}$
is renormalizable but not gauge invariant, despite the fact that the
object $W^{\alpha}A_{\alpha}$ is both renormalizable and gauge invariant.
In this case the gauge invariance is destroyed by the presence of
a number of gauge superfields larger than two in the vertex. It is
important to point out that the gauge symmetry in this renormalization
analysis must be imposed by hand since the superfield perturbation
formalism is not manifestly gauge invariant.

The scalar and gauge-scalar sectors are particularly interesting from
the renormalization point of view. Note first that the renormalization
condition ($N_{\Phi}+N_{D}\leq4$) implies that any operator constructed
with more than four scalar superfields is immediately non-renormalizable.
Hence $N_{\Phi}$ may solely take three values: $N_{\Phi}=0,\,2,\,4$.
Since the $N_{\Phi}=0$ case (gauge sector) has been already studied,
we are going to proceed with the $N_{\Phi}=2$ case. Here any renormalizable
operator must contain at most two susy-covariant derivatives ($N_{D}\leq2$).
Thus as workable operators one has
\begin{equation}
\bar{\Phi}\Phi,\quad D^{\alpha}\bar{\Phi}D_{\alpha}\Phi,\quad A^{\alpha}D_{\alpha}\bar{\Phi}\Phi,\quad A^{\alpha}\bar{\Phi}D_{\alpha}\Phi,\quad A^{\alpha}A_{\alpha}\bar{\Phi}\Phi,\quad\cdots\label{eq:2.10}
\end{equation}
where the ellipsis stands for other operators with $N_{D}+N_{A}>2$.
Note that the object $D^{\alpha}A_{\alpha}\bar{\Phi}\Phi$ is not
included above because this is a linear combination of the third and
fourth operators after integration by parts. Similarly $D^{\alpha}\bar{\Phi}D_{\alpha}\Phi$
is equivalent by by-part integration to $\bar{\Phi}D^{2}\Phi$. 

Making use of the gauge transformations (\ref{eq:2.3}) we may easily
verify that the first operator in (\ref{eq:2.10}), i.e. $\bar{\Phi}\Phi$,
is gauge invariant, while the others are not (at least independently).
Note however that since the transformation of the gauge field $A_{\alpha}$
involves a susy-covariant derivative, we might expect that the linear
combination of the four remaining operators in (\ref{eq:2.10}) turns
out to be gauge invariant. After a dimensional analysis of the operators
in (\ref{eq:2.10}) and invoking the Hermiticity property of the action
as a whole, it is easy to verify that such linear combination has
to be of the form
\begin{equation}
D^{\alpha}\bar{\Phi}D_{\alpha}\Phi+ieu\left(A^{\alpha}D_{\alpha}\bar{\Phi}\Phi-A^{\alpha}\bar{\Phi}D_{\alpha}\Phi\right)+e^{2}vA^{2}\bar{\Phi}\Phi,\label{eq:2.11}
\end{equation}
where $A^{2}\doteq A^{\alpha}A_{\alpha}/2$ and $u$ and $v$ are
dimensionless constants. Requiring the invariance of this expression
under the gauge transformations (\ref{eq:2.3}), we straightforwardly
identify that $u=-1$ and $v=2$. 

By introducing the object $\nabla_{\alpha}\doteq D_{\alpha}-ieA_{\alpha}$,
we can encapsulate the four operators in (\ref{eq:2.11}) into a single
operator. That this,
\begin{equation}
\bar{\nabla}^{\alpha}\bar{\Phi}\nabla_{\alpha}=D^{\alpha}\bar{\Phi}D_{\alpha}\Phi-ie\left(A^{\alpha}D_{\alpha}\bar{\Phi}\Phi-A^{\alpha}\bar{\Phi}D_{\alpha}\Phi\right)+2e^{2}A^{2}\bar{\Phi}\Phi.\label{eq:2.12}
\end{equation}
The object $\nabla_{\alpha}$ is called gauge covariant derivative
due to its transformation property under (\ref{eq:2.3}): $\nabla_{\alpha}\rightarrow\mbox{e}^{ieK}\nabla_{\alpha}\mbox{e}^{-ieK}$.
In terms of this gauge-covariant derivative we can easily construct
other gauge invariant operators. However, as we have seen, the only
operator which satisfies all the physical requirements is that shown
in (\ref{eq:2.12}). Note that any other operator with more than two
covariant derivatives $\nabla_{\alpha}$ turns out non-renormalizable.
Finally, for $N_{\Phi}=4$, which in turn implies that $N_{D}=0$,
one finds the scalar self-coupling $\left(\bar{\Phi}\Phi\right)^{2}$
as the only physically acceptable operator.

In sum, we have shown that the most general renormalizable supersymmetric
gauge theory is described by the action
\begin{equation}
S=\int d^{5}z\left\{ \frac{1}{2}W^{\alpha}W_{\alpha}+m\, W^{\alpha}A_{\alpha}-\frac{1}{2}\bar{\nabla}^{\alpha}\bar{\Phi}\nabla_{\alpha}\Phi+M\,\bar{\Phi}\Phi+\lambda\left(\bar{\Phi}\Phi\right)^{2}\right\} .\label{eq:2.13}
\end{equation}
This action (with lower-derivative operators) is the supersymmetric
version of the usual Maxwell-Chern-Simons theory coupled to matter
superfield.

From now on, we confine ourselves to study the ultraviolet behavior
of the massless supersymmetric quantum electrodynamics (SQED$_{3}$).
In particular, we study the possibility of removing the residual susy
divergences in the two-loop scalar and gauge self-energies at two-loop
level by introducing appropriate higher derivative operators in the
kinetic part of the action. 

Setting $m=0=M$ in (\ref{eq:2.13}) and using the gauge fixing term
(\ref{eq:2.6}), the action for the SQED$_{3}$ with scalar self-coupling
reads
\begin{equation}
S=\int d^{5}z\left\{ \frac{1}{2}W^{\alpha}W_{\alpha}-\frac{1}{2}\bar{\nabla}^{\alpha}\bar{\Phi}\nabla_{\alpha}\Phi+\lambda\left(\bar{\Phi}\Phi\right)^{2}-\frac{1}{4\alpha}D^{\alpha}A_{\alpha}D^{2}D^{\beta}A_{\beta}+\bar{C}D^{2}C\right\} .\label{eq:2.14}
\end{equation}
Here $C$ and $\bar{C}$ are respectively the ghost and anti-ghost
superfields. This action is invariant under the following BRST transformations
\begin{equation}
\delta\Phi=ie\Lambda C\Phi,\quad\delta A_{\alpha}=-\Lambda D_{\alpha}C,\quad\delta C=0,\quad\delta\bar{C}=\frac{1}{\alpha}\Lambda D^{2}D^{\alpha}A_{\alpha},
\end{equation}
where $\Lambda$ is a Grassmann parameter. Notice that as in usual
field theories these transformations are nilpotent, i. e. $\delta^{2}=0$.
On the other hand, since the ghost and anti-ghost do not couple with
the other fields, they can be ignored in the perturbative analysis.

Using the propagators in (\ref{eq:2.7a}-\ref{eq:2.7b}) and identifying
all kinds of vertices in the action (\ref{eq:2.14}), 
\begin{equation}
S_{int}=\int d^{5}z\left\{ \frac{ie}{2}\left(A^{\alpha}D_{\alpha}\bar{\Phi}\Phi-A^{\alpha}\bar{\Phi}D_{\alpha}\Phi\right)-e^{2}A^{2}\bar{\Phi}\Phi+\lambda\left(\bar{\Phi}\Phi\right)^{2}\right\} ,\label{eq:2.15}
\end{equation}
we compute easily the superficial degree of divergence for the SQED$_{3}$,
\begin{equation}
\omega=2-\frac{1}{2}V_{1}^{\left(3\right)}-V_{0}^{\left(4\right)}-\frac{1}{2}E_{\Phi}-\frac{1}{2}n_{D},\label{eq:2.16}
\end{equation}
where $V_{N_{D}}^{(N_{l})}$ denotes the number of scalar-gauge vertices
with $N_{l}$ lines and $N_{D}$ susy-covariant derivatives ($D_{\alpha}$),
while $E_{\Phi}$ and $n_{D}$ have the same meanings as in (\ref{eq:2.8a}).

\begin{figure}
\begin{centering}
\includegraphics[scale=0.7]{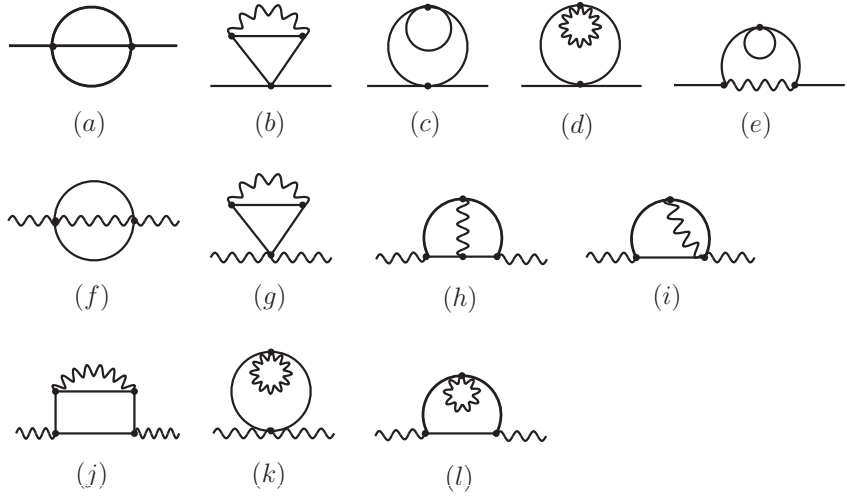}
\par\end{centering}

\caption{\label{fig:SQED3}Divergent radiative corrections to the scalar and
gauge self-energies for the SQED$_{3}$ with scalar self-interaction.
Solid lines represent scalar propagators, while wavy lines represent
gauge propagators.}
\end{figure}

If we rewrite (\ref{eq:2.16}) in the way $\omega=2-x$, with $x$
representing all the other terms, and spotting that $x$ is strictly
an integer positive number ($x\geq1$), it is easy to realize that
the (lower-derivative) SQED$_{3}$ with scalar self-coupling possesses
just logarithmic $\left(\omega=0\right)$ and linear $\left(\omega=1\right)$
divergences. These divergences ($1/\epsilon$) will appear from the
two-loop diagrams, for in planar physics and within the dimensional
reduction (DReD$_{3}$) scheme \cite{Siegel (1979-1980)} all one-loop
diagrams are finite \cite{Avdeev-etal (1992-3)}. In Fig. \ref{fig:SQED3}
we display all two-loop contributions to the self-energy of the scalar
and gauge superfields which are divergent on power counting grounds.
It is important to point out however that diagrams (c), (d), (e),
(k) and (l) in Fig. \ref{fig:SQED3} turn out to be finite after performing
the D-algebra. The reason is that each two-loop momentum integral
can be expressed as the product of two independent finite one-loop
integrals. The remaining infinities in the scalar and gauge self-energies
at two-loop level must be removed by the usual renormalization procedure.
In this respect, it has been shown in \cite{Lehum etal (2008)} that
there is an unusual gauge in which the SQED$_{3}$ (without scalar
self-interaction) turns out finite. In the next section we introduce
appropriate higher-derivative operators in the standard SQED$_{3}$
action to remove these infinities and so improve the ultraviolet behavior
of the usual theory.

\section{Higher-derivative SQED$_{3}$ and two-loop finiteness of the self-energies
\label{sec:Sec3}}

This section is devoted to implement two higher derivative operators
in the lower-derivative SQED$_{3}$ with scalar self-coupling. These
operators are judiciously chosen and inserted into the kinetic part
of the action for improving the asymptotic behavior of the propagators,
so that to get rid of all two-loop divergences in the scalar and gauge
self-energies. Clearly, according to the preceding discussion, these
sorts of operators are non-renormalizable in the usual theory.

Since the field strength $W_{\alpha}$ is invariant ($W'_{\alpha}=W_{\alpha}$)
under gauge transformations, there is an infinity of possible higher
derivative operators in the gauge sector which obey all symmetries
of the standard theory. Nevertheless there is only one operator which
removes partly the divergences, but that is the supersymmetric extension
of the operator $\partial_{\mu}F^{\mu\nu}\partial_{\rho}F_{\;\nu}^{\rho}$
which epitomizes the four dimensional Lee-Wick quantum electrodynamics
(Lee-Wick-QED$_{4}$). It is not hard to see that this higher derivative
operator is given by $D^{2}W^{\alpha}D^{2}W_{\alpha}$. Indeed after
performing the integral over $\theta$ it exhibits the following content
\begin{equation}
\left.D^{2}\left(D^{2}W^{\alpha}D^{2}W_{\alpha}\right)\right|=\frac{1}{2}\partial_{\mu}F^{\mu\nu}\partial_{\rho}F_{\;\nu}^{\rho}+2\lambda^{\alpha}i\partial_{\alpha}^{\;\;\beta}\square\lambda_{\beta},\label{eq:3.1}
\end{equation}
where the spinor field $\lambda_{\alpha}$ is the susy partner (photino)
of the photon $v^{\mu}$ field.

On the other hand, an examination of the superficial degree of divergence
of the theory, which results of adding this operator in (\ref{eq:2.14}),
reveals that it is necessary to introduce one more operator in this
scalar sector to remove successfully all divergences in the diagrams
displayed in Fig. \ref{fig:SQED3}. Needless to say, this additional
operator must be constructed by using the gauge covariant derivative
$\nabla_{\alpha}=D_{\alpha}-iA_{\alpha}$ in order to respect the
gauge symmetry. This higher derivative operator becomes to be $\bar{\nabla}^{2}\bar{\Phi}\nabla^{2}\Phi$,
where $\nabla^{2}\doteq\nabla^{\alpha}\nabla_{\alpha}/2$, and gives
rise to new interaction vertices with five and six lines on incorporating
in the standard SQED$_{3}$. The higher-derivative vertices are given
by 
\begin{eqnarray}
S'_{int} & = & b\int d^{5}z\left(\bar{\nabla}^{2}\bar{\Phi}\nabla^{2}\Phi-D^{2}\bar{\Phi}D^{2}\Phi\right)=b\int d^{5}z\left[\frac{ie}{2}\left(\bar{\Sigma}D^{2}\Phi-\Sigma D^{2}\bar{\Phi}\right)\right.\nonumber \\
 &  & \left.-e^{2}A^{2}\left(\bar{\Phi}D^{2}\Phi+\Phi D^{2}\bar{\Phi}\right)+\frac{ie^{3}}{2}A^{2}\left(\bar{\Phi}\Sigma-\Phi\bar{\Sigma}\right)+\frac{e^{2}}{4}\bar{\Sigma}\Sigma+e^{4}A^{4}\bar{\Phi}\Phi\right].\label{eq:3.2}
\end{eqnarray}
Here for simplicity we have introduced the superfield $\Sigma\left(A,\,\Phi\right)\doteq D^{\alpha}\left(A_{\alpha}\Phi\right)+A^{\alpha}D_{\alpha}\Phi$.

Taking into account these two operators, our higher-derivative supersymmetric
quantum electrodynamics in three dimensions (HSQED$_{3}$) is defined
by
\begin{eqnarray}
S & = & \int d^{5}z\left\{ \frac{1}{2}W^{\alpha}W_{\alpha}+\frac{a}{2}\, D^{2}W^{\alpha}D^{2}W_{\alpha}-\frac{1}{2}\bar{\nabla}^{\alpha}\bar{\Phi}\nabla_{\alpha}\Phi+b\bar{\nabla}^{2}\bar{\Phi}\nabla^{2}\Phi+\lambda\left(\bar{\Phi}\Phi\right)^{2}\right\} \label{eq:3.3}
\end{eqnarray}
Note that the negative dimensions of the higher derivative coefficients
($\left[a\right]=-2$, $\left[b\right]=-1$) mirror the non-renormalizability
of their corresponding operators, as discussed in the foregoing section.

It is interesting to see how these higher derivatives are distributed
among the component fields. For simplicity let us consider merely
the quadratic part of the Lagrangian. Carrying out the $\theta$-integral
in (\ref{eq:3.3}), using the $\theta$-Taylor expansions (\ref{eq:2.1}-\ref{eq:2.2}),
it is easy to show that the quadratic Lagrangian is given by 
\begin{eqnarray}
\mathcal{L}_{0} & = & -\frac{1}{8}F^{\mu\nu}F_{\mu\nu}+\lambda i\partial\lambda+\frac{a}{4}\partial_{\mu}F^{\mu\nu}\partial_{\rho}F_{\nu}^{\rho}+a\lambda i\partial\square\lambda+\bar{\varphi}\square\varphi+\bar{\psi}i\partial\psi\nonumber \\
 &  & +\bar{F}F+b\,\left(\bar{\varphi}\square F+\varphi\square\bar{F}\right)+b\bar{\psi}\square\psi.
\end{eqnarray}
Here the spinor indices are contracted by following the north-west
rule $\left(\searrow\right)$ and the square of a spinor includes
a factor of 1/2 in its definition. Notice that the scalar field $\varphi$
acquires a higher-derivative operator only in the on-shell formulation,
i.e. after eliminating the auxiliary field $F$ through its equation
of motion $\left(F=-b\square\varphi\right)$.

These higher-derivative operators do not modified the classical potential.
This may be seen by setting $\Phi=\varphi-\theta^{2}F$ and $A_{\alpha}=0$
into (\ref{eq:3.3}), with $\varphi$ and $F$ labeling constants.
Doing this and ignoring an over-all space-time integral $\int d^{3}x$
, we get for the classical potential
\begin{equation}
V_{cl}=-\int d^{2}\theta\left\{ \bar{\Phi}D^{2}\Phi+\lambda\left(\bar{\Phi}\Phi\right)^{2}\right\} =-\bar{F}F-2\lambda\bar{\varphi}\varphi\left(\bar{\varphi}F+\varphi\bar{F}\right).
\end{equation}
If we eliminate the auxiliary field $F$, writing $\varphi=\frac{1}{\sqrt{2}}\left(\varphi_{r}+i\varphi_{i}\right)$,
the classical potential can be written as $V_{cl}=\frac{\lambda^{2}}{2}\left(\varphi_{r}^{2}+\varphi_{i}^{2}\right)^{3}$.
Thus at classical level supersymmetry remains intact. At this point
it is worth mentioning that the Kählerian potential at one-loop level
was recently calculated for a family of three-dimensional superfield
Abelian gauge theories with higher-derivative operators in the gauge
sector \cite{Gama-Petrov (2013)}. 

In what follows we shall show that the higher-derivative kinetic operators
introduced in the standard SQED$_{3}$ improve the asymptotic behavior
of the propagators in such a way that the scalar and gauge self-energies
at two-loop order become to be finite.

The propagators for the scalar and gauge fields associated with the
HSQED$_{3}$ action are given by
\begin{equation}
\left\langle \bar{\Phi}\left(k,\,\theta\right)\Phi\left(-k,\,\theta\right)\right\rangle =-i\frac{\left(D^{2}+b\, k^{2}\right)}{k^{2}\left(1+b^{2}k^{2}\right)}\delta^{2}\left(\theta-\theta'\right)\label{eq:3.4a}
\end{equation}
and 
\begin{equation}
\left\langle A_{\alpha}\left(k,\,\theta\right)A_{\beta}\left(-k,\,\theta\right)\right\rangle =\frac{i}{2\left(k^{2}\right)^{2}}\left[\frac{1}{\left(1-ak^{2}\right)}D_{\beta}D_{\alpha}-\alpha D_{\alpha}D_{\beta}\right]D^{2}\delta^{2}\left(\theta-\theta'\right).\label{eq:3.4b}
\end{equation}
Here once again we have fixed the gauge through (\ref{eq:2.6}). The
asymptotic behavior of these propagators by virtue of the relation
$\left\{ D_{\alpha},\, D_{\beta}\right\} =2k_{\alpha\beta}$ is
\begin{equation}
\left\langle \bar{\Phi}\Phi\right\rangle \sim\frac{1}{b^{2}k^{3}}+\frac{1}{bk^{2}}\qquad\qquad\left\langle AA\right\rangle \sim\frac{1}{ak^{4}}-\frac{\alpha}{k^{2}}\sim\frac{1}{k^{4}}\,\,\,\mbox{(Landau gauge)}.\label{eq:3.5}
\end{equation}
Note that these propagators are more convergent than their counterparts
(\ref{eq:2.7a}-\ref{eq:2.7b}), whichever value for the gauge parameter
$\alpha$ is chosen.

To work out the superficial degree of divergence for the HSQED$_{3}$,
we regard the less convergent part (i.e. $1/k^{2}$) of the scalar
propagator and choose the Landau gauge ($\alpha=0$) in order to obtain
a well-defined asymptotic gauge propagator. In this way we get 
\begin{equation}
\omega=2-2P_{A}-\frac{3}{2}V_{1}^{\left(3\right)}-\frac{1}{2}V_{3}^{\left(3\right)}-2V_{0}^{\left(4\right)}-V_{2}^{\left(4\right)}-2V_{0}^{\Phi}-\frac{3}{2}V_{1}^{\left(5\right)}-2V_{0}^{\left(6\right)}-\frac{1}{2}n_{D}\label{eq:3.6}
\end{equation}
where $V_{0}^{\Phi}$ and $P_{A}$ denote the number of pure scalar
vertices and the number of gauge propagators, respectively. The other
variables were defined in (\ref{eq:2.16}). An immediate consequence
of this result is that any Feynman diagrams with one or more internal
gauge propagators turns out to be convergent on power counting grounds.

\begin{figure}[h]
\begin{centering}
\includegraphics[scale=0.6]{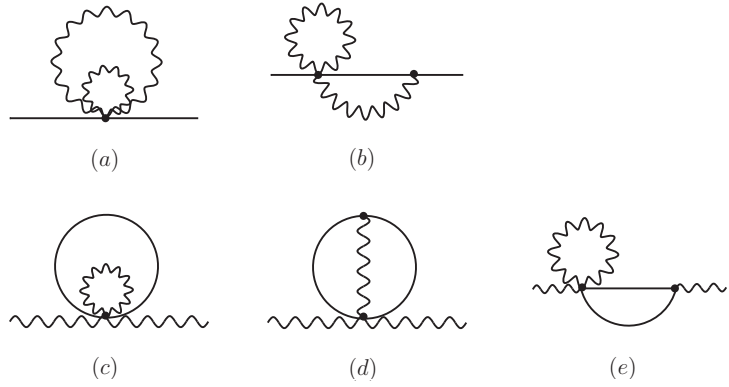}
\par\end{centering}

\caption{\label{fig:HSQED3}Radiative corrections to the scalar and gauge self-energies
due to the higher derivative vertices.}
\end{figure}

A simple analysis of the superficial degree of divergence (\ref{eq:3.6})
shows that all Feynman diagrams in Fig. \ref{fig:SQED3}, which are
divergent within the standard SQED$_{3}$, become now finite. In this
case there are however other contributions to the scalar and gauge
self-energies that we must regard. These two-loop contributions correspond
to the extra higher-derivative interactions (\ref{eq:3.2}) and are
depicted in Fig. \ref{fig:HSQED3}. By invoking once again the power
counting criterion, it is easy to show that these extra diagrams are
also convergent. In this way we have shown that the introduction of
two higher-derivative operators in the kinetic part of the SQED$_{3}$
action improves the ultraviolet behavior of the theory, allowing in
particular the elimination of all two-loop divergences in the scalar
and gauge self-energies.

The key role of the ghosts in removing UV divergences is hidden in
the super-power counting analysis. In the component formalism the
identification of the ghosts is relatively simple. They become evident
either by reformulating the higher-derivative theory in terms of lower-derivative
operators (a theory with indefinite metric), i.e. by eliminating the
higher-derivative operators by means of auxiliary fields (ghosts),
or by splitting the higher-derivative propagators, using the expression
\begin{equation}
\frac{1}{\left(k^{2}+m_{1}^{2}\right)\left(k^{2}+m_{2}^{2}\right)}=\frac{1}{m_{2}^{2}-m_{1}^{2}}\left[\frac{1}{k^{2}+m_{1}^{2}}-\frac{1}{k^{2}+m_{2}^{2}}\right],\label{eq:3.7}
\end{equation}
into elementary propagators, where the propagators with ``wrong sign''
correspond to the ghosts. In the superfield formalism, by contrast,
the ghosts become visible only after performing the Grassmann reduction
procedure (D-algebra) in the Feynman amplitudes so that the formula
(\ref{eq:3.7}) can be used to identify the ghost contributions. In
our previous work \cite{Gall-Sen-Silv (2013)} we have shown, by evaluating
explicitly the diagram (a) in Fig. \ref{fig:SQED3}, how the mutual
cancellation between ghost divergences and ``normal'' ones happens.
Note that in this kind of theory the ``normal'' divergences are
indeed the residual divergences of the susy mechanism of removing
UV infinities. Finally, it is important to mention that the two-loop
momentum integrals in \cite{Gall-Sen-Silv (2013)} were calculated
by using the master integral \cite{Tan-etal} {\small{
\begin{equation}
\int_{k,q}\frac{1}{\left(k^{2}+x^{2}\right)\left(q^{2}+y^{2}\right)\left[\left(k+q\right)^{2}+z^{2}\right]}=\frac{\mu^{-2\epsilon}}{32\pi^{2}}\left\{ \frac{1}{\epsilon}-\gamma+1-\ln\left[\frac{\left(x+y+z\right)^{2}}{4\pi\mu^{2}}\right]\right\} +\mathcal{B}_{res}\label{eq:3.8}
\end{equation}
where}} $\epsilon=3-D$, with $D$ denoting the spacetime dimension,
$\gamma$ is the Euler's constant, and $\mathcal{B}_{res}$ stands
for the total residue contribution of the complex poles inside an
energy contour appropriate for performing the ``Wick-rotation''.
As commented in the introduction, in the ultraviolet analysis, we
are assuming that the residue contributions are finite.

\section{Conclusions\label{sec:Sec4}}

Within the $\mathcal{N}=1$ superfield formalism, we have deformed
the standard supersymmetric quantum electrodynamics in three dimensions
(SQED$_{3}$) by introducing two higher-derivative operators in its
kinetic action. One of them is the susy extension of the higher-derivative
operator $\partial_{\mu}F^{\mu\nu}\partial_{\rho}F_{\;\nu}^{\rho}$
in Lee-Wick-QED in four dimensions and the other is the gauge extension
of the operator $D^{2}\bar{\Phi}D^{2}\Phi$ in the HWZ$_{3}$ model.
These operators respect all of the symmetries of the original model,
but are non-renormalizable on power counting grounds. In the on-shell
component formulation, these operators introduce higher derivatives
for all component fields. Here the auxiliary field $F$ plays the
role of a Lee-Wick field by introducing the higher-derivative operator
for the the scalar field $\varphi$ only in the on-shell case. At
the quantum level, we show, under a rather general assumption about
the complex poles, that the ghosts associated with these two higher-derivative
operators cancel out all residual susy divergences in the scalar and
gauge self-energies at two-loop level. 
\begin{acknowledgments}
This work has been supported by the CAPES-Brazil.\end{acknowledgments}

\end{document}